\documentclass[journal=jacsat,manuscript=article]{achemso}

\usepackage[version=3]{mhchem} 
\usepackage[T1]{fontenc} 

\usepackage{color}
\usepackage{xspace}
\usepackage{geometry}
\usepackage{graphicx}
\usepackage{float}
\usepackage{chemfig}
\usepackage{subcaption}
\usepackage{adjustbox}
\usepackage{amsfonts}
\usepackage{etoolbox}
\usepackage{hyperref}
\usepackage{pgf}

\usepackage[normalem]{ulem} 

\pgfkeys{
  /molecules/1/.initial=\ce{MPF}\xspace,
  /molecules/3/.initial=\ce{CCl3{-}MPF}\xspace,
  /molecules/4/.initial=\ce{Cl{-}MPF}\xspace,
  /molecules/5/.initial=\ce{$4$\textsc{'}Cl{-}MPF}\xspace,
  /molecules/7/.initial=\ce{CCl3{-}$4$\textsc{'}Cl{-}MPF}\xspace,
  /molecules/8/.initial=\ce{Cl{-}$4$\textsc{'}Cl{-}MPF}\xspace,
  /molecules/9/.initial=\ce{FMT}\xspace,
  /molecules/11/.initial=\ce{CCl3{-}FMT}\xspace,
  /molecules/12/.initial=\ce{Cl{-}FMT}\xspace,
  /molecules/13/.initial=\ce{FYT}\xspace,
  /molecules/15/.initial=\ce{CCl3{-}FYT}\xspace,
  /molecules/16/.initial=\ce{Cl{-}FYT}\xspace,
  /molecules/17/.initial=\ce{MMPIF}\xspace,
  /molecules/19/.initial=\ce{CCl3{-}MMPIF}\xspace,
  /molecules/20/.initial=\ce{Cl{-}MMPIF}\xspace,
  /molecules/25/.initial=\ce{TIF}\xspace,
  /molecules/27/.initial=\ce{CCl3{-}TIF}\xspace,
  /molecules/28/.initial=\ce{Cl{-}TIF}\xspace
}

\newcommand{\molecule}[1]{\pgfkeysvalueof{/molecules/#1}}

\author{Ivan Tambovtsev}
\affiliation[University of Iceland]{Science Institute and Faculty of Physical Sciences, University of Iceland, 107 Reykjav\'{\i}k, Iceland}
\email{ivt3@hi.is}
\author{\'Oskar Kristinsson}
\affiliation[University of Iceland]{Science Institute and Faculty of Physical Sciences, University of Iceland, 107 Reykjav\'{\i}k, Iceland}
\email{oskarkristinsson@mailbox.org}
\author{Hannes J\'onsson}
\affiliation[University of Iceland]{Science Institute and Faculty of Physical Sciences, University of Iceland, 107 Reykjav\'{\i}k, Iceland}
\email{hj@hi.is}

\title{The Effect of Chlorine Substitution on Rotational Speed and Light Absorption of Second Generation Molecular Motors
}

\begin{document}

\renewcommand*\tocentryname{TOC Graphic}
\begin{tocentry}
    \includegraphics[width = \textwidth]{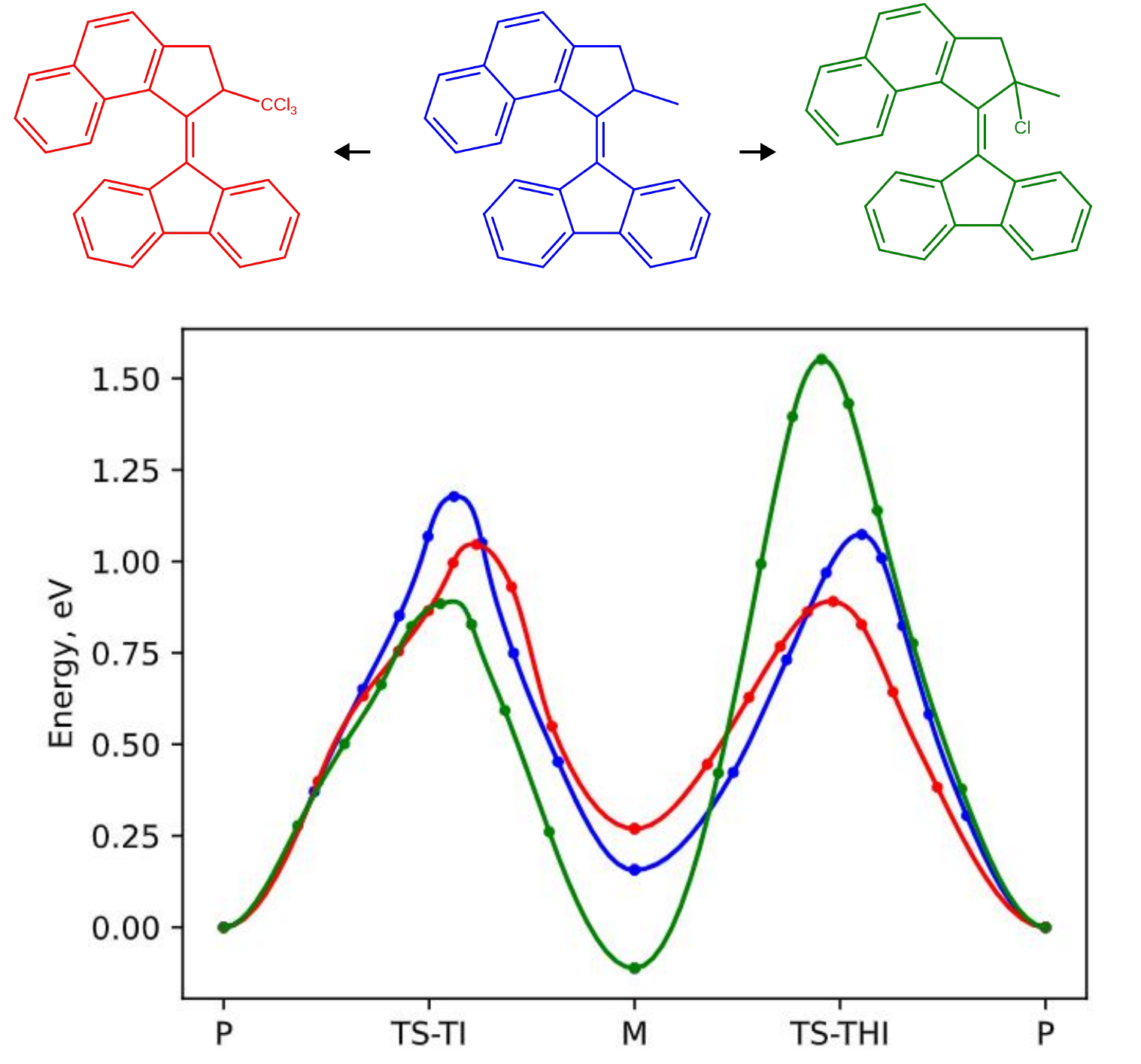}
\end{tocentry}

\begin{abstract}
The effect of substituting a hydrogen atom by a chlorine atom or a methyl group by a trichloromethyl ($\ce{CCl3}$) group at the stereogenic center of light-driven second generation molecular motors is calculated in order to assess the effect on rotational speed and the separation of the absorption peaks of the isomers. While experimental and theoretical studies have previously been carried out for fluorine substitution, this is the first study of chlorine substitution. Five well-characterized base molecules are studied and the trends are compared with the effect of fluorine substitution. The trichloromethyl substitution is found to accelerate the rotation more than a trifluoromethyl ($\ce{CF3}$) substitution by reducing the life-time of the metastable state, due to larger steric hindrance in the metastable state than in the transition state for the thermal helix inversion (THI). A larger increase in the separation of the absorption peaks of the two isomers is also obtained. The Cl atom substitution, however, changes the energy landscape significantly, making the M isomer lower in energy than the P isomer, and raising the energy barrier for THI beyond that of the back transition, thus quenching the rotation.
\end{abstract}


\section*{Introduction}

Nature’s molecular motors, refined over millions of years of evolution, provide inspiration for the design of synthetic molecular machines, and this has been an active field of research in the past couple of decades.~\cite{feringaArtBuildingSmall2017} 
Possible applications are in 
various fields, such as optics, photonics and light-driven soft materials. 
Second generation molecular motors are characterized by a central C=C double bond that connects a ``rotor" part of the molecule to a ``stator" part and rotates half a circle for each photon absorbed. 
In the conventional operating cycle~\cite{rokeMolecularRotaryMotors2018}, the molecule starts from the more stable isomer,
denoted as
P,~\cite{goldIUPACCompendiumChemical2019, koumuraSecondGenerationLightDriven2002}
and, upon absorption of a photon, undergoes a ca. 90$^\circ$ rotation in the excited state followed by a further rotation in the ground state to the higher-energy, metastable isomer, denoted as M.  
This photogenerated state is the crucial intermediate from which two competing thermally activated transitions determine the molecule's ultimate function. If the molecule undergoes a conformational change corresponding to completion of the first half of the rotation, the so-called thermal helix inversion (THI), a repeat of these steps after absorption of a second photon completes a full cycle of the motor. If, however, the preferred thermally activated transition from the M state is a backward transition to the original isomer, the so-called thermal isomerization (TI), the molecule does not rotate, but could function as a thermally reversible (T-type) molecular switch~\cite{iriePhotochromismDiaryletheneMolecules2014}. The ultimate performance and role of the molecule is, therefore, affected by the competition between the THI and TI transitions in the ground electronic state.

Optimization of the rotational speed is a critical aspect of molecular motor design, as the ideal frequency is highly application-dependent. High speeds are sought for soft actuators and drug delivery systems that rely on rapid mechanical work~\cite{guinartSyntheticMolecularMotor2023, houPhotoresponsiveHelicalMotion2021, rajonsonOptimizingMotionFolding2018}, 
whereas slow rotation is required for applications such as catalysis, where a specific isomeric state 
may be used
for stereochemical control~\cite{wangDynamicControlChiral2011}. In other systems, function emerges from a precisely controlled dynamic interplay, as seen in the formation of revolving structures in liquid crystals~\cite{orlovaRevolvingSupramolecularChiral2018}. For second-generation motors, a wide variety of designs with different rotational speeds have been developed to meet these diverse demands~\cite{poolerDesigningLightdrivenRotary2021}. The speed can be altered through major changes in the molecular geometry, such as replacing a six-membered ring in the rotor with a five-membered one~\cite{vicarioControllingSpeedRotation2005}, but one way of fine-tuning that only weakly affects the molecular structure is, for example, fluorine substitution~\cite{huangLongLivedSupramolecularHelices2018, blegerOFluoroazobenzenesReadilySynthesized2012, stackoFluorineSubstitutedMolecularMotors2017, tambovtsevFineTuningRotational2025}.

\begin{figure}[H]
    \centering
    \includegraphics[width=0.7\textwidth]{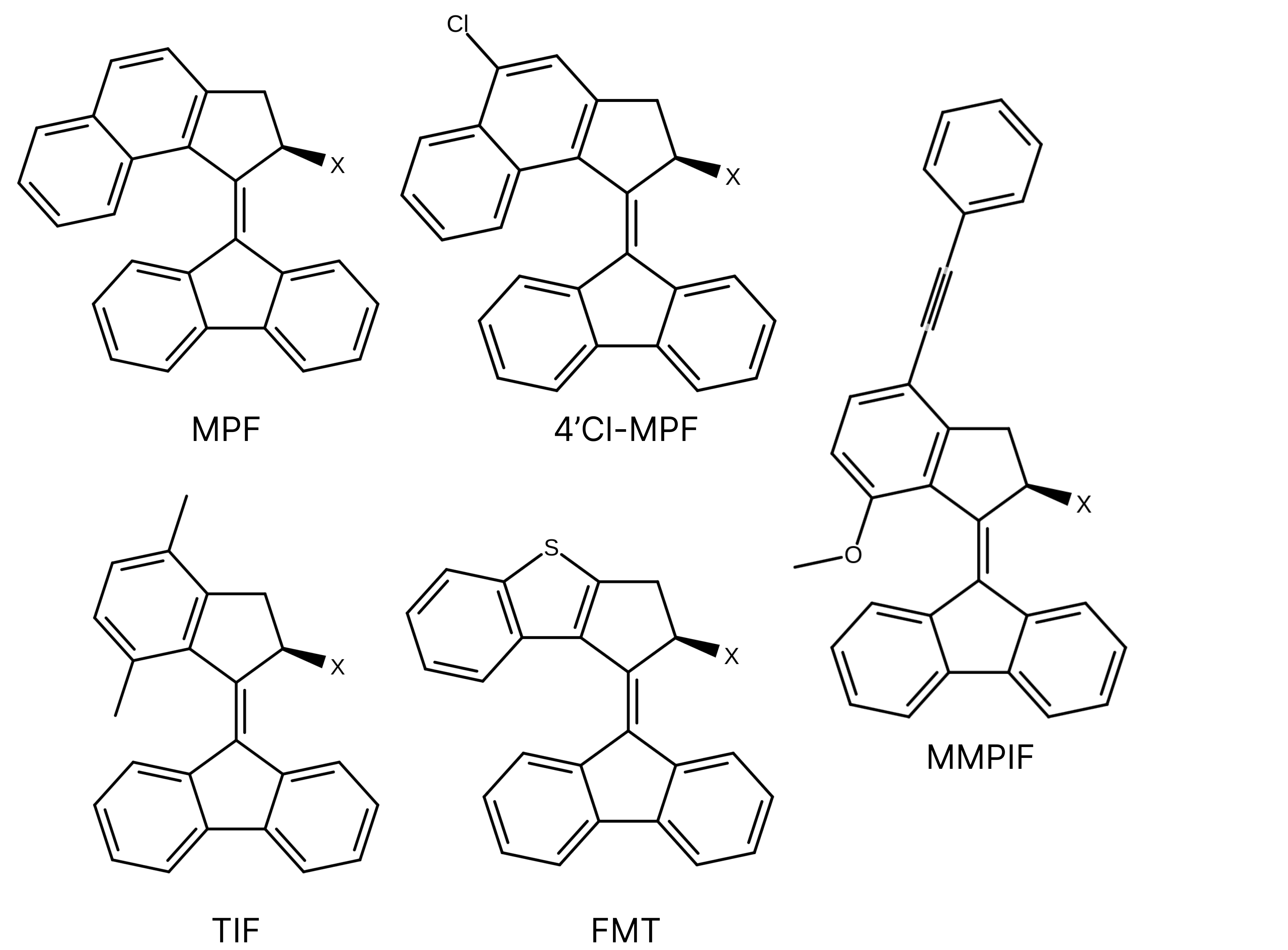}
    \caption{
    The second-generation molecular motors studied here with substitutional site marked with an X.
    Two modifications are studied: H\,$\to$\,Cl and Me\,$\to$\,\ce{CCl3}.
    }
    \label{fig:molecules}
\end{figure}

Here, we study the effect chlorination at the stereogenic center can have on the properties of the set of five second-generation molecular motors shown in Figure~\ref{fig:molecules}. 
These molecules are representative of the common second-generation motor designs, featuring a fluorene stator and a five-membered ring in the rotor connected by a central C=C double bond. They have been previously synthesized and experimentally studied~\cite{vicarioFineTuningRotary2006, pollardRedesignLightdrivenRotary2008, pollardEffectDonorAcceptor2008, cnossenTrimerUltrafastNanomotors2009, garcia-lopezLightActivatedOrganicMolecular2020}, and were chosen for this study because they exhibit a diverse range of structural features and rotational properties, providing a robust basis for evaluating the effect of modifications in a systematic way. 
The effect of replacing a methyl group at site X with a trichloromethyl (\ce{CCl3}) group is calculated
as well as the effect of replacing a hydrogen atom at this site with a chlorine atom. 
The results are compared with previously reported effects of fluorine substitutions.~\cite{tambovtsevFineTuningRotational2025}
The separation between the absorption peaks of the two isomers is also calculated since a large enough gap is important for selectively driving the rotation.


\section*{Results and Discussion}
\begin{figure}[H]
\centering
\includegraphics[width = 0.9\textwidth]{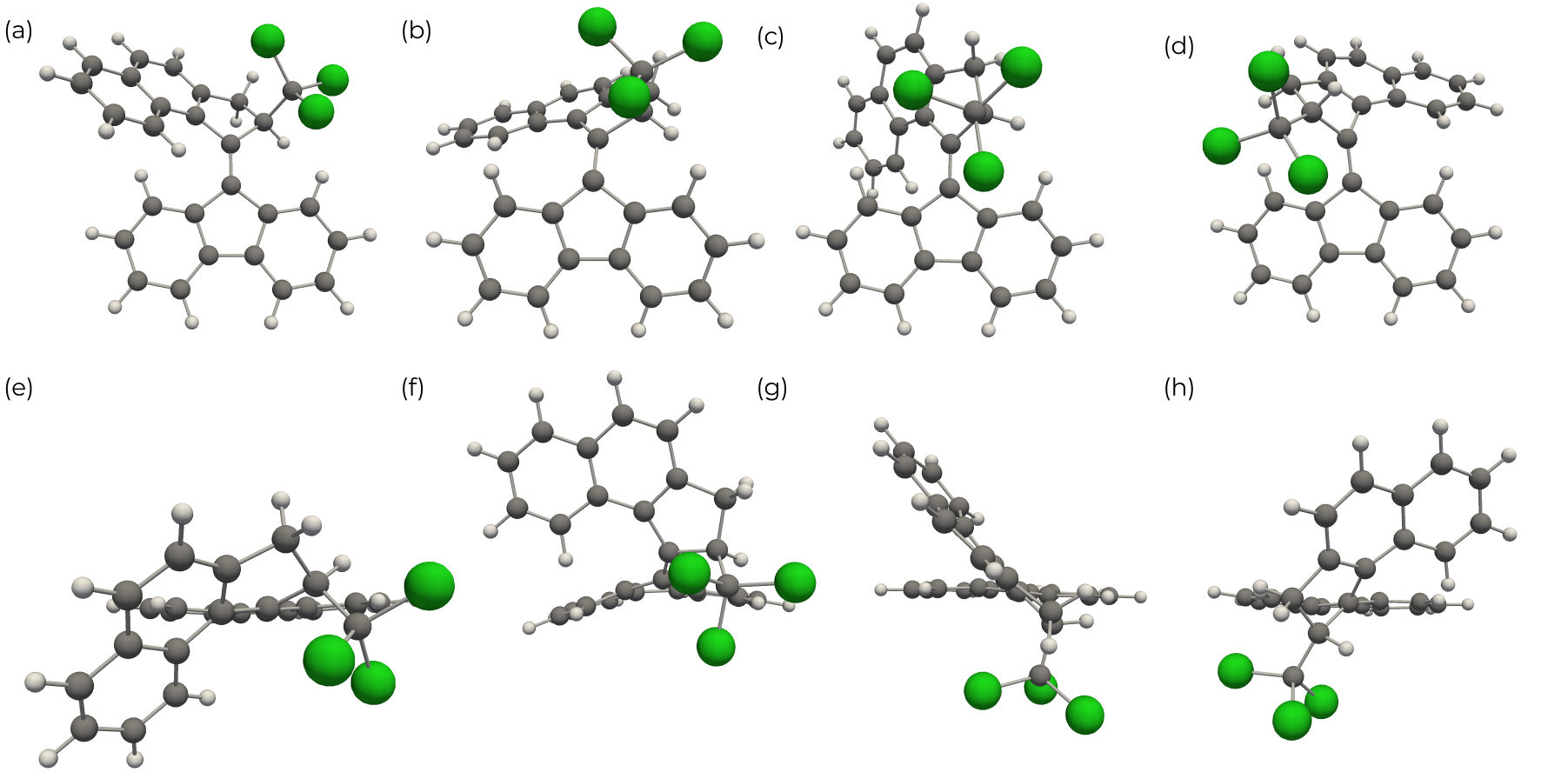}
\caption{
Front and top views of the optimized structures for the various states of molecule \molecule{3}. 
(a and e) The stable P isomer.
(b and f) The transition structure for the forward thermal helix inversion, THI.
(c and g) The metastable M isomer.
(d and h) The transition structure for the backward thermal isomerization, TI. 
}
\label{fig:structures}
\end{figure}

 Figure~\ref{fig:structures} shows the stable, P, and metastable, M, isomers as well as the THI and TI transition structures (corresponding to first order saddle points on the energy surface) for molecule \molecule{3}. 
The steric interactions at the stereogenic center, in particular at the M isomer and the two transition structures, are critical in determining the rotational speed as discussed below.

\begin{figure}[H]
\centering
\includegraphics[width = \textwidth]{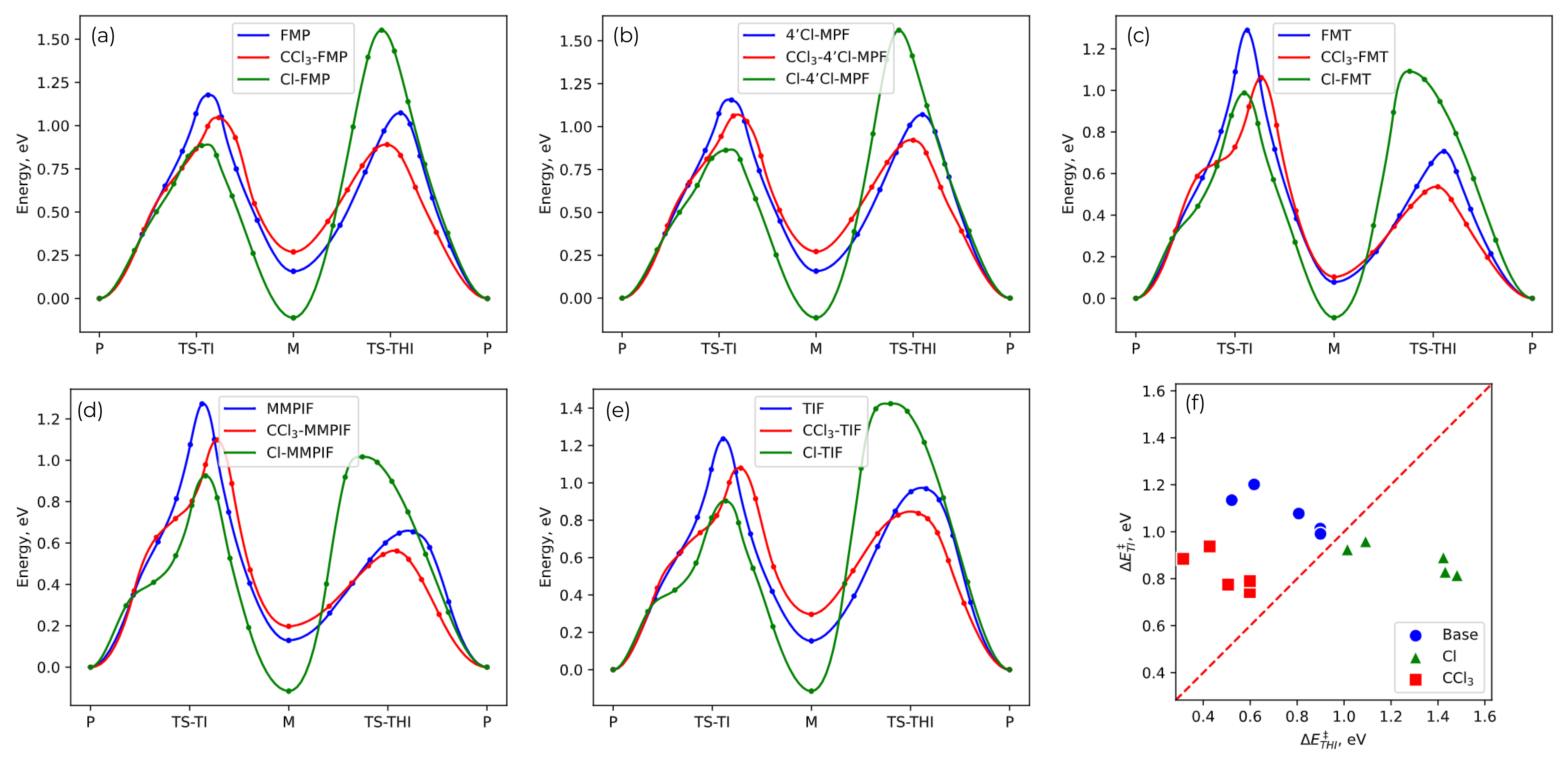}
\caption{
(a-e) Minimum energy paths between the P and M isomers of the five base molecules where a methyl 
group and a hydrogen atom are at the X site (blue), as well as the chlorinated molecules with the methyl group substituted for a \ce{CCl3} group (red), or the H atom is substituted for a Cl atom (green).
The zero of energy is taken to be the P state in each case. 
The energy maxima correspond to first order saddle points, i.e.  transition structures, for the TI (left) and THI (right) transitions.
For the \ce{CCl3} substitution, both energy barriers decrease and the THI barrier continues to be lower than that for TI. 
For the Cl substitution, the M isomer becomes more stable than P, and the THI energy barrier increases significantly. 
(f) Activation energy for the TI and THI transitions from the metastable state (the M isomer for the base (blue) and \ce{CCl3} substituted (red) molecules, but the P state for Cl substituted (green) molecules). The dashed red line separates molecules by their dominant thermal transition mechanism: THI for those above the line and TI for those below.
}
\label{fig:mep}
\end{figure}

\begin{figure}[H]
\centering
\includegraphics[width = 0.45\textwidth]{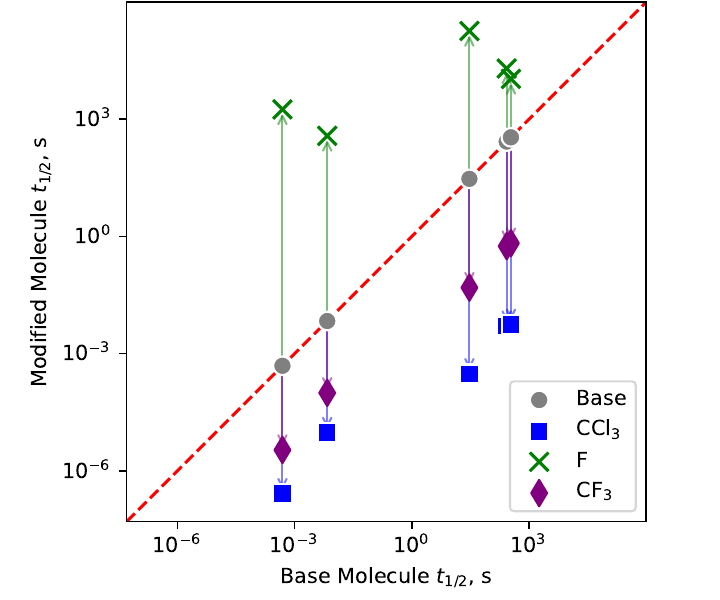}
\caption{Comparison of the half-life of the metastable M isomer of the molecules. The substitution of the methyl group for a $\ce{CCl3}$ group (blue) shortens the half-life and thereby increases the rotational speed, more so than the previously reported substitution by a $\ce{CF3}$ group. Data on fluorine and $\ce{CF3}$ substitution are from Ref.~\citenum{tambovtsevFineTuningRotational2025}.
}
\label{fig:fit}
\end{figure}

Figure~\ref{fig:mep} shows the
calculated minimum energy paths (MEPs) for the THI and TI transitions of the five base molecules as well as the chlorine substituted variants.
The replacement of the methyl group by a \ce{CCl3} group consistently lowers the energy barrier for both transitions.
The calculated rate constants are changed accordingly, the one for THI increasing by two to five orders of magnitude (see 
Table~\ref{table:Spectroscopy}).
This translates to a significant shortening of the lifetime of the metastable state and a corresponding increase in the rotational speed.
In all cases, the forward THI continues to have a lower energy barrier than TI so the molecules still function as molecular motors. 
Figure~\ref{fig:fit} illustrates the change in the calculated half-life of the metastable state.

\begin{table}[H]
\centering
\caption{
Absorption wavelengths (in nm) for the two isomers and calculated rate constants (in s$^{-1}$) for the forward (THI) and backward (TI) thermal transitions at 298 K. 
Note that the rate constants are for transitions from the higher energy, metastable isomer in each case.
For the lower block of the table, the Cl substituted molecules, this is the P state (see figure 3).
}
\label{table:Spectroscopy}
\begin{tabular}{|c|c|c|c|c|}
\hline
Molecule & \textbf{$\lambda_{\text{P}}$} & \textbf{$\lambda_{\text{M}}$} & $k^{\rm THI}$ & $k^{\rm TI}$ \\
\hline
\molecule{1} & 379 & 410 & $2.58 \times 10^{-3}$ & $8.23 \times 10^{-4}$ \\
\molecule{3} & 381 & 436 & $1.38 \times 10^{2}$ & $7.07 \times 10^{0}$ \\
\molecule{5} & 388 & 420 & $2.01 \times 10^{-3}$ & $1.74 \times 10^{-3}$ \\
\molecule{7} & 386 & 443 & $1.24 \times 10^{2}$ & $5.38 \times 10^{-2}$ \\
\molecule{9} & 347 & 365 & $1.01 \times 10^{2}$ & $1.13 \times 10^{-6}$ \\
\molecule{11} & 362 & 393 & $7.28 \times 10^{4}$ & $5.85 \times 10^{-3}$ \\
\molecule{17} & 349 & 389 & $1.42 \times 10^{3}$ & $8.78 \times 10^{-6}$ \\
\molecule{19} & 347 & 426 & $2.67 \times 10^{6}$ & $3.68 \times 10^{-2}$ \\
\molecule{25} & 340 & 367 & $2.32 \times 10^{-2}$ & $7.93 \times 10^{-7}$ \\
\molecule{27} & 346 & 398 & $2.28 \times 10^{3}$ & $3.53 \times 10^{0}$ \\
\hline
\hline
\molecule{4} & 399 & 416 & $1.18 \times 10^{-12}$ & $5.57 \times 10^{-1}$ \\
\molecule{8} & 408 & 422 & $3.55 \times 10^{-13}$ & $9.21 \times 10^{-1}$ \\
\molecule{12} & 361 & 385 & $1.73 \times 10^{-6}$ & $7.28 \times 10^{-3}$ \\
\molecule{20} & 365 & 376 & $3.76 \times 10^{-5}$ & $2.94 \times 10^{-2}$ \\
\molecule{28} & 359 & 378 & $1.24 \times 10^{-12}$ & $1.09 \times 10^{-3}$ \\
\hline
\end{tabular}
\end{table}

In contrast, the substitution of a hydrogen atom at the stereogenic center with a chlorine atom 
is found to produce a qualitatively different outcome. The MEPs shown in figure 3 change significantly, resulting in a reversal of the relative stability of the two isomers. 
The M isomer becomes the stable one, while the P isomer becomes metastable. This stability reversal is accompanied by a large increase in the energy barrier for the THI. 
As a result, the dominant thermal relaxation pathway is no longer the forward rotation step for the Cl substituted molecules. 
A similar but smaller increase in the THI energy barrier was observed for fluorine substitution making the TI in some cases faster than THI.~\cite{stackoFluorineSubstitutedMolecularMotors2017,tambovtsevFineTuningRotational2025}

Figure~\ref{fig:fit} illustrates the calculated half-life of the  metastable state of the molecules that can function as molecular motors (with Me, F, \ce{CF3}, or \ce{CCl3} at the X site). The half-life is dominated by the rate of the forward THI step. The shorter the half-life is, the faster the rotation. The \ce{CCl3} substitution leads to a dramatic shortening of the half-life, 
in a more pronounced way than was previously reported for \ce{CF3} substitution~\cite{tambovtsevFineTuningRotational2025}. 
The calculated rate constants, listed in Table~\ref{table:Spectroscopy}, show that the forward THI process continues to be the dominant thermal pathway, faster than TI, and thus leading to unidirectional rotation.

\begin{figure}[H]
  \centering
  \includegraphics[width = 0.45\textwidth]{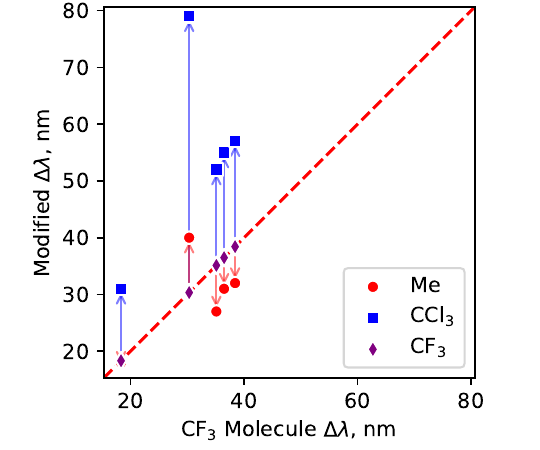}
  \caption{
  Difference in excitation wavelength, $\Delta \lambda$, of the 
  two isomers of the molecules with methyl (red), \ce{CCl3} (blue) and \ce{CF3} (purple)~\cite{tambovtsevFineTuningRotational2025} groups at the stereogenic center.
  The substitution of the methyl for \ce{CCl3} consistently increases the gap and thereby would improve selectivity in the photoexcitation.
  }
  \label{fig:spectroscopy}
\end{figure}

Figure~\ref{fig:spectroscopy} illustrates the difference in the calculated absorption wavelengths for the two isomers. 
A clear and advantageous trend emerges for the $\ce{CCl3}$ substitution. For all the molecules, the \ce{CCl3} substitution consistently gives the largest gap between the absorption peaks of the two isomers. This large spectral separation is a significant practical advantage, as it makes it easier to selectively drive the motor cycle by photoactivation.

The H to Cl substitution leads to more drastic changes.  The stability of the isomers is inverted, but still the stable isomer absorbs at a longer wavelength than the metastable isomer (see Table~\ref{table:Spectroscopy}). Spectroscopically, this might be favorable as it would in principle allow for the selective excitation of the stable isomer with lower-energy light. However, it is not clear what the outcome of such an excitation would be. The absorption of a photon does not guarantee efficient isomerization to the metastable state; the molecule could follow other, non-productive excited-state relaxation pathways. Therefore, to determine whether these altered molecules can function as rotors or as photoswitches, a study of their excited-state dynamics is required to understand the photochemical reaction path and quantum yield.

\section*{Conclusion}

In summary, the effect of chlorination of second generation molecular motors is explored by theoretical calculations. 
The results show that a methyl to \ce{CCl3} substitution at the stereographic center will increase the speed of rotation and also improve spectroscopic properties for selective activation, more so than the previously studied \ce{CF3} substitution.
However, the substitution of the hydrogen atom at the same center by a \ce{Cl} atom leads to a more drastic change in the energy landscape, reversing the relative stability of the two isomers, P and M, and increasing the energy barrier for the THI transition beyond that of the TI transition. Further studies of the photochemical quantum yield for the Cl-substituted molecules are needed to reveal the dominant photorelaxation mechanism, necessary to determine whether the 
H$\to$Cl 
replacement 
can give
a molecular switch or a motor. 



\section{Methods}

The minimum energy paths (MEPs) for the THI and TI transitions were calculated using the climbing image nudged elastic band (CI-NEB) method~\cite{millsReversibleWorkTransition1995,henkelmanClimbingImageNudged2000,henkelmanImprovedTangentEstimate2000},
using energy-weighted springs~\cite{asgeirssonNudgedElasticBand2021}
as implemented in the ORCA software~\cite{neeseSoftwareUpdateORCA2025},.
The initial path was generated using the sequential image dependent pair potential (S-IDPP) method~\cite{schmerwitzImprovedInitializationOptimal2024}.
The position of each image on the path was converged to a tolerance of $2.5\cdot 10^{-3}$\,a.u.\ (RMS) and $5\cdot 10^{-3}$\,a.u.\ (MAX) for the force perpendicular to the local tangent, with one order of magnitude tighter tolerance for the climbing image. The climbing image was then made to converge fully on the first order saddle point using a saddle point search method~\cite{asgeirssonNudgedElasticBand2021} until the atomic forces had dropped below $10^{-4}$\,a.u.\ (RMS) and $3\cdot 10^{-4}$\,a.u.\ (MAX).

The DFT calculations were carried out using the B3LYP hybrid functional approximation\cite{leeDevelopmentColleSalvettiCorrelationenergy1988, beckeDensityfunctionalExchangeenergyApproximation1988, beckeDensityFunctionalThermochemistry1993} 
and a linear combination of atomic orbitals formalism employing the 6-31G(d,p) basis set~\cite{hehreSelfConsistentMolecular2003, weigendAccurateCoulombfittingBasis2006}.
The B3LYP functional is chosen for its proven performance in a wide range of molecular systems, and the 6-31G(d,p) basis set is selected for its balance between computational efficiency and accuracy.

The ground state electronic structure calculations were converged to thresholds of $10^{-7}$\,a.u.\ for the maximum component of the density change, $5\cdot 10^{-9}$\,a.u.\ for the root mean square (RMS) of the density change, and $5\cdot 10^{-7}$\,a.u.\ for the error in the direct inversion in the iterative subspace (DIIS). The structure of the stable and metastable states of all molecular motors were optimized to a tolerance of $10^{-4}$\,a.u.\ (RMS) and $3\cdot 10^{-4}$\,a.u.\ (MAX) for the gradient, and $2\cdot 10^{-3}$\,a.u.\ (RMS) and $4\cdot 10^{-3}$\,a.u.\ (MAX) for the optimization step. 

The rate constant for thermally activated transitions was calculated using the harmonic approximation to transition state theory (HTST)~\cite{wignerTransitionStateMethod1938,vineyardFrequencyFactorsIsotope1957},
\begin{equation}
k_{\rm HTST} = \frac{\prod_i^{3N} \nu_i^{\rm min}}{\prod_i^{3N-1} \nu_i^\ddagger}
\exp \left[ -\frac{E^\ddagger - E^{\rm min}}{k_{\rm B} T} \right],
\label{eqn:htst-rate}
\end{equation}
where $\nu_i^{\rm min}$ and $\nu_i^\ddagger$ refer to vibrational frequencies at the minimum and first-order saddle point, respectively, and $E^{\rm min}$ and $E^\ddagger$ denote the corresponding values of the energy (zero point energy corrections included). 
The overall half-life of the metastable isomers was calculated as $t_{1/2}=\ln 2/(k_{\rm THI}+k_{\rm TI})$.
The methodology described above for estimating the half-life of the metastable isomer has previously been tested and found to give 
results in close agreement with experimental measurements for a wide range of molecular motors.~\cite{tambovtsevFineTuningRotational2025}

The spectra were calculated using linear-response TDDFT within the adiabatic approximation. All calculations were performed with the ORCA 6.1 software~\cite{neeseSoftwareUpdateORCA2025}.
The data was extracted using ChemParse.

\begin{acknowledgement}
This work was funded by the Icelandic Research Fund (grants 239970 and 2511544). 
We thank Tetiana Orlova and Gianluca Levi for helpful discussions.
The calculations were carried out at the IREI HPC facility at the University of Iceland.

\end{acknowledgement}

\section{Data Availability Statement}
The data supporting the findings of this work are available for download at Zenodo.


\bibliography{zotero_tambovtsev}

\providecommand{\latin}[1]{#1}
\makeatletter
\providecommand{\doi}
  {\begingroup\let\do\@makeother\dospecials
  \catcode`\{=1 \catcode`\}=2 \doi@aux}
\providecommand{\doi@aux}[1]{\endgroup\texttt{#1}}
\makeatother
\providecommand*\mcitethebibliography{\thebibliography}
\csname @ifundefined\endcsname{endmcitethebibliography}  {\let\endmcitethebibliography\endthebibliography}{}
\begin{mcitethebibliography}{35}
\providecommand*\natexlab[1]{#1}
\providecommand*\mciteSetBstSublistMode[1]{}
\providecommand*\mciteSetBstMaxWidthForm[2]{}
\providecommand*\mciteBstWouldAddEndPuncttrue
  {\def\EndOfBibitem{\unskip.}}
\providecommand*\mciteBstWouldAddEndPunctfalse
  {\let\EndOfBibitem\relax}
\providecommand*\mciteSetBstMidEndSepPunct[3]{}
\providecommand*\mciteSetBstSublistLabelBeginEnd[3]{}
\providecommand*\EndOfBibitem{}
\mciteSetBstSublistMode{f}
\mciteSetBstMaxWidthForm{subitem}{(\alph{mcitesubitemcount})}
\mciteSetBstSublistLabelBeginEnd
  {\mcitemaxwidthsubitemform\space}
  {\relax}
  {\relax}

\bibitem[Feringa(2017)]{feringaArtBuildingSmall2017}
Feringa,~B.~L. The {{Art}} of {{Building Small}}: {{From Molecular Switches}} to {{Motors}} ({{Nobel Lecture}}). \emph{Angewandte Chemie International Edition} \textbf{2017}, \emph{56}, 11060--11078\relax
\mciteBstWouldAddEndPuncttrue
\mciteSetBstMidEndSepPunct{\mcitedefaultmidpunct}
{\mcitedefaultendpunct}{\mcitedefaultseppunct}\relax
\EndOfBibitem
\bibitem[Roke \latin{et~al.}(2018)Roke, Wezenberg, and Feringa]{rokeMolecularRotaryMotors2018}
Roke,~D.; Wezenberg,~S.~J.; Feringa,~B.~L. Molecular Rotary Motors: {{Unidirectional}} Motion around Double Bonds. \emph{Proceedings of the National Academy of Sciences of the United States of America} \textbf{2018}, \emph{115}, 9423--9431\relax
\mciteBstWouldAddEndPuncttrue
\mciteSetBstMidEndSepPunct{\mcitedefaultmidpunct}
{\mcitedefaultendpunct}{\mcitedefaultseppunct}\relax
\EndOfBibitem
\bibitem[Gold(2019)]{goldIUPACCompendiumChemical2019}
Gold,~V., Ed. \emph{The {{IUPAC Compendium}} of {{Chemical Terminology}}: {{The Gold Book}}}, 4th ed.; {International Union of Pure and Applied Chemistry (IUPAC)}: Research Triangle Park, NC, 2019\relax
\mciteBstWouldAddEndPuncttrue
\mciteSetBstMidEndSepPunct{\mcitedefaultmidpunct}
{\mcitedefaultendpunct}{\mcitedefaultseppunct}\relax
\EndOfBibitem
\bibitem[Koumura \latin{et~al.}(2002)Koumura, Geertsema, {van Gelder}, Meetsma, and Feringa]{koumuraSecondGenerationLightDriven2002}
Koumura,~N.; Geertsema,~E.~M.; {van Gelder},~M.~B.; Meetsma,~A.; Feringa,~B.~L. Second {{Generation Light-Driven Molecular Motors}}. {{Unidirectional Rotation Controlled}} by a {{Single Stereogenic Center}} with {{Near-Perfect Photoequilibria}} and {{Acceleration}} of the {{Speed}} of {{Rotation}} by {{Structural Modification}}. \emph{Journal of the American Chemical Society} \textbf{2002}, \emph{124}, 5037--5051\relax
\mciteBstWouldAddEndPuncttrue
\mciteSetBstMidEndSepPunct{\mcitedefaultmidpunct}
{\mcitedefaultendpunct}{\mcitedefaultseppunct}\relax
\EndOfBibitem
\bibitem[Irie \latin{et~al.}(2014)Irie, Fukaminato, Matsuda, and Kobatake]{iriePhotochromismDiaryletheneMolecules2014}
Irie,~M.; Fukaminato,~T.; Matsuda,~K.; Kobatake,~S. Photochromism of {{Diarylethene Molecules}} and {{Crystals}}: {{Memories}}, {{Switches}}, and {{Actuators}}. \emph{Chemical Reviews} \textbf{2014}, \emph{114}, 12174--12277\relax
\mciteBstWouldAddEndPuncttrue
\mciteSetBstMidEndSepPunct{\mcitedefaultmidpunct}
{\mcitedefaultendpunct}{\mcitedefaultseppunct}\relax
\EndOfBibitem
\bibitem[Guinart \latin{et~al.}(2023)Guinart, Korpidou, Doellerer, Pacella, Stuart, Dinu, Portale, Palivan, and Feringa]{guinartSyntheticMolecularMotor2023}
Guinart,~A.; Korpidou,~M.; Doellerer,~D.; Pacella,~G.; Stuart,~M. C.~A.; Dinu,~I.~A.; Portale,~G.; Palivan,~C.; Feringa,~B.~L. Synthetic Molecular Motor Activates Drug Delivery from Polymersomes. \emph{Proceedings of the National Academy of Sciences} \textbf{2023}, \emph{120}, e2301279120\relax
\mciteBstWouldAddEndPuncttrue
\mciteSetBstMidEndSepPunct{\mcitedefaultmidpunct}
{\mcitedefaultendpunct}{\mcitedefaultseppunct}\relax
\EndOfBibitem
\bibitem[Hou \latin{et~al.}(2021)Hou, Mondal, Long, {de Haan}, Zhao, Zhou, Liu, Broer, Chen, and Feringa]{houPhotoresponsiveHelicalMotion2021}
Hou,~J.; Mondal,~A.; Long,~G.; {de Haan},~L.; Zhao,~W.; Zhou,~G.; Liu,~D.; Broer,~D.~J.; Chen,~J.; Feringa,~B.~L. Photo-Responsive {{Helical Motion}} by {{Light-Driven Molecular Motors}} in a {{Liquid-Crystal Network}}. \emph{Angewandte Chemie International Edition} \textbf{2021}, \emph{60}, 8251--8257\relax
\mciteBstWouldAddEndPuncttrue
\mciteSetBstMidEndSepPunct{\mcitedefaultmidpunct}
{\mcitedefaultendpunct}{\mcitedefaultseppunct}\relax
\EndOfBibitem
\bibitem[Rajonson \latin{et~al.}(2018)Rajonson, Ciobotarescu, and Teboul]{rajonsonOptimizingMotionFolding2018}
Rajonson,~G.; Ciobotarescu,~S.; Teboul,~V. Optimizing the Motion of a Folding Molecular Motor in Soft Matter. \emph{Physical Chemistry Chemical Physics} \textbf{2018}, \emph{20}, 10077--10085\relax
\mciteBstWouldAddEndPuncttrue
\mciteSetBstMidEndSepPunct{\mcitedefaultmidpunct}
{\mcitedefaultendpunct}{\mcitedefaultseppunct}\relax
\EndOfBibitem
\bibitem[Wang and Feringa(2011)Wang, and Feringa]{wangDynamicControlChiral2011}
Wang,~J.; Feringa,~B.~L. Dynamic {{Control}} of {{Chiral Space}} in a {{Catalytic Asymmetric Reaction Using}} a {{Molecular Motor}}. \emph{Science} \textbf{2011}, \emph{331}, 1429--1432\relax
\mciteBstWouldAddEndPuncttrue
\mciteSetBstMidEndSepPunct{\mcitedefaultmidpunct}
{\mcitedefaultendpunct}{\mcitedefaultseppunct}\relax
\EndOfBibitem
\bibitem[Orlova \latin{et~al.}(2018)Orlova, Lancia, Loussert, Iamsaard, Katsonis, and Brasselet]{orlovaRevolvingSupramolecularChiral2018}
Orlova,~T.; Lancia,~F.; Loussert,~C.; Iamsaard,~S.; Katsonis,~N.; Brasselet,~E. Revolving Supramolecular Chiral Structures Powered by Light in Nanomotor-Doped Liquid Crystals. \emph{Nature Nanotechnology} \textbf{2018}, \emph{13}, 304--308\relax
\mciteBstWouldAddEndPuncttrue
\mciteSetBstMidEndSepPunct{\mcitedefaultmidpunct}
{\mcitedefaultendpunct}{\mcitedefaultseppunct}\relax
\EndOfBibitem
\bibitem[Pooler \latin{et~al.}(2021)Pooler, Lubbe, Crespi, and Feringa]{poolerDesigningLightdrivenRotary2021}
Pooler,~D. R.~S.; Lubbe,~A.~S.; Crespi,~S.; Feringa,~B.~L. Designing Light-Driven Rotary Molecular Motors. \emph{Chemical Science} \textbf{2021}, \emph{12}, 14964--14986\relax
\mciteBstWouldAddEndPuncttrue
\mciteSetBstMidEndSepPunct{\mcitedefaultmidpunct}
{\mcitedefaultendpunct}{\mcitedefaultseppunct}\relax
\EndOfBibitem
\bibitem[Vicario \latin{et~al.}(2005)Vicario, Meetsma, and Feringa]{vicarioControllingSpeedRotation2005}
Vicario,~J.; Meetsma,~A.; Feringa,~B.~L. Controlling the Speed of Rotation in Molecular Motors. {{Dramatic}} Acceleration of the Rotary Motion by Structural Modification. \emph{Chemical Communications} \textbf{2005}, 5910--5912\relax
\mciteBstWouldAddEndPuncttrue
\mciteSetBstMidEndSepPunct{\mcitedefaultmidpunct}
{\mcitedefaultendpunct}{\mcitedefaultseppunct}\relax
\EndOfBibitem
\bibitem[Huang \latin{et~al.}(2018)Huang, Orlova, Matt, and Katsonis]{huangLongLivedSupramolecularHelices2018}
Huang,~H.; Orlova,~T.; Matt,~B.; Katsonis,~N. Long-{{Lived Supramolecular Helices Promoted}} by {{Fluorinated Photoswitches}}. \emph{Macromolecular Rapid Communications} \textbf{2018}, \emph{39}, 1700387\relax
\mciteBstWouldAddEndPuncttrue
\mciteSetBstMidEndSepPunct{\mcitedefaultmidpunct}
{\mcitedefaultendpunct}{\mcitedefaultseppunct}\relax
\EndOfBibitem
\bibitem[Bléger \latin{et~al.}(2012)Bléger, Schwarz, Brouwer, and Hecht]{blegerOFluoroazobenzenesReadilySynthesized2012}
Bléger,~D.; Schwarz,~J.; Brouwer,~A.~M.; Hecht,~S. O-{{Fluoroazobenzenes}} as {{Readily Synthesized Photoswitches Offering Nearly Quantitative Two-Way Isomerization}} with {{Visible Light}}. \emph{Journal of the American Chemical Society} \textbf{2012}, \emph{134}, 20597--20600\relax
\mciteBstWouldAddEndPuncttrue
\mciteSetBstMidEndSepPunct{\mcitedefaultmidpunct}
{\mcitedefaultendpunct}{\mcitedefaultseppunct}\relax
\EndOfBibitem
\bibitem[Štacko \latin{et~al.}(2017)Štacko, Kistemaker, and Feringa]{stackoFluorineSubstitutedMolecularMotors2017}
Štacko,~P.; Kistemaker,~J. C.~M.; Feringa,~B.~L. Fluorine-{{Substituted Molecular Motors}} with a {{Quaternary Stereogenic Center}}. \emph{Chemistry – A European Journal} \textbf{2017}, \emph{23}, 6643--6653\relax
\mciteBstWouldAddEndPuncttrue
\mciteSetBstMidEndSepPunct{\mcitedefaultmidpunct}
{\mcitedefaultendpunct}{\mcitedefaultseppunct}\relax
\EndOfBibitem
\bibitem[Tambovtsev \latin{et~al.}(2025)Tambovtsev, Schmerwitz, Levi, Darmoroz, Nesterov, Orlova, and Jónsson]{tambovtsevFineTuningRotational2025}
Tambovtsev,~I.; Schmerwitz,~Y. L.~A.; Levi,~G.; Darmoroz,~D.~D.; Nesterov,~P.~V.; Orlova,~T.; Jónsson,~H. Fine {{Tuning}} of the {{Rotational Speed}} of {{Light-Driven}}, {{Second-Generation Molecular Motors}} by {{Fluorine Substitution}}. \emph{The Journal of Physical Chemistry Letters} \textbf{2025}, 4014--4020\relax
\mciteBstWouldAddEndPuncttrue
\mciteSetBstMidEndSepPunct{\mcitedefaultmidpunct}
{\mcitedefaultendpunct}{\mcitedefaultseppunct}\relax
\EndOfBibitem
\bibitem[Vicario \latin{et~al.}(2006)Vicario, Walko, Meetsma, and Feringa]{vicarioFineTuningRotary2006}
Vicario,~J.; Walko,~M.; Meetsma,~A.; Feringa,~B.~L. Fine {{Tuning}} of the {{Rotary Motion}} by {{Structural Modification}} in {{Light-Driven Unidirectional Molecular Motors}}. \emph{Journal of the American Chemical Society} \textbf{2006}, \emph{128}, 5127--5135\relax
\mciteBstWouldAddEndPuncttrue
\mciteSetBstMidEndSepPunct{\mcitedefaultmidpunct}
{\mcitedefaultendpunct}{\mcitedefaultseppunct}\relax
\EndOfBibitem
\bibitem[Pollard \latin{et~al.}(2008)Pollard, Meetsma, and Feringa]{pollardRedesignLightdrivenRotary2008}
Pollard,~M.~M.; Meetsma,~A.; Feringa,~B.~L. A Redesign of Light-Driven Rotary Molecular Motors. \emph{Organic \& Biomolecular Chemistry} \textbf{2008}, \emph{6}, 507--512\relax
\mciteBstWouldAddEndPuncttrue
\mciteSetBstMidEndSepPunct{\mcitedefaultmidpunct}
{\mcitedefaultendpunct}{\mcitedefaultseppunct}\relax
\EndOfBibitem
\bibitem[Pollard \latin{et~al.}(2008)Pollard, Wesenhagen, Pijper, and Feringa]{pollardEffectDonorAcceptor2008}
Pollard,~M.~M.; Wesenhagen,~P.~V.; Pijper,~D.; Feringa,~B.~L. On the Effect of Donor and Acceptor Substituents on the Behaviour of Light-Driven Rotary Molecular Motors. \emph{Organic \& Biomolecular Chemistry} \textbf{2008}, \emph{6}, 1605--1612\relax
\mciteBstWouldAddEndPuncttrue
\mciteSetBstMidEndSepPunct{\mcitedefaultmidpunct}
{\mcitedefaultendpunct}{\mcitedefaultseppunct}\relax
\EndOfBibitem
\bibitem[Cnossen \latin{et~al.}(2009)Cnossen, Pijper, Kudernac, Pollard, Katsonis, and Feringa]{cnossenTrimerUltrafastNanomotors2009}
Cnossen,~A.; Pijper,~D.; Kudernac,~T.; Pollard,~M.~M.; Katsonis,~N.; Feringa,~B.~L. A Trimer of Ultrafast Nanomotors: Synthesis, Photochemistry and Self-Assembly on Graphite. \emph{Chemistry (Weinheim an Der Bergstrasse, Germany)} \textbf{2009}, \emph{15}, 2768--2772\relax
\mciteBstWouldAddEndPuncttrue
\mciteSetBstMidEndSepPunct{\mcitedefaultmidpunct}
{\mcitedefaultendpunct}{\mcitedefaultseppunct}\relax
\EndOfBibitem
\bibitem[{García-López} \latin{et~al.}(2020){García-López}, Liu, and Tour]{garcia-lopezLightActivatedOrganicMolecular2020}
{García-López},~V.; Liu,~D.; Tour,~J.~M. Light-{{Activated Organic Molecular Motors}} and {{Their Applications}}. \emph{Chemical Reviews} \textbf{2020}, \emph{120}, 79--124\relax
\mciteBstWouldAddEndPuncttrue
\mciteSetBstMidEndSepPunct{\mcitedefaultmidpunct}
{\mcitedefaultendpunct}{\mcitedefaultseppunct}\relax
\EndOfBibitem
\bibitem[Mills \latin{et~al.}(1995)Mills, Jónsson, and Schenter]{millsReversibleWorkTransition1995}
Mills,~G.; Jónsson,~H.; Schenter,~G.~K. Reversible Work Transition State Theory: Application to Dissociative Adsorption of Hydrogen. \emph{Surface Science} \textbf{1995}, \emph{324}, 305--337\relax
\mciteBstWouldAddEndPuncttrue
\mciteSetBstMidEndSepPunct{\mcitedefaultmidpunct}
{\mcitedefaultendpunct}{\mcitedefaultseppunct}\relax
\EndOfBibitem
\bibitem[Henkelman \latin{et~al.}(2000)Henkelman, Uberuaga, and Jónsson]{henkelmanClimbingImageNudged2000}
Henkelman,~G.; Uberuaga,~B.~P.; Jónsson,~H. A Climbing Image Nudged Elastic Band Method for Finding Saddle Points and Minimum Energy Paths. \emph{The Journal of Chemical Physics} \textbf{2000}, \emph{113}, 9901--9904\relax
\mciteBstWouldAddEndPuncttrue
\mciteSetBstMidEndSepPunct{\mcitedefaultmidpunct}
{\mcitedefaultendpunct}{\mcitedefaultseppunct}\relax
\EndOfBibitem
\bibitem[Henkelman and Jónsson(2000)Henkelman, and Jónsson]{henkelmanImprovedTangentEstimate2000}
Henkelman,~G.; Jónsson,~H. Improved Tangent Estimate in the Nudged Elastic Band Method for Finding Minimum Energy Paths and Saddle Points. \emph{The Journal of Chemical Physics} \textbf{2000}, \emph{113}, 9978--9985\relax
\mciteBstWouldAddEndPuncttrue
\mciteSetBstMidEndSepPunct{\mcitedefaultmidpunct}
{\mcitedefaultendpunct}{\mcitedefaultseppunct}\relax
\EndOfBibitem
\bibitem[Ásgeirsson \latin{et~al.}(2021)Ásgeirsson, Birgisson, Bjornsson, Becker, Neese, Riplinger, and Jónsson]{asgeirssonNudgedElasticBand2021}
Ásgeirsson,~V.; Birgisson,~B.~O.; Bjornsson,~R.; Becker,~U.; Neese,~F.; Riplinger,~C.; Jónsson,~H. Nudged {{Elastic Band Method}} for {{Molecular Reactions Using Energy-Weighted Springs Combined}} with {{Eigenvector Following}}. \emph{Journal of Chemical Theory and Computation} \textbf{2021}, \emph{17}, 4929--4945\relax
\mciteBstWouldAddEndPuncttrue
\mciteSetBstMidEndSepPunct{\mcitedefaultmidpunct}
{\mcitedefaultendpunct}{\mcitedefaultseppunct}\relax
\EndOfBibitem
\bibitem[Neese(2025)]{neeseSoftwareUpdateORCA2025}
Neese,~F. Software {{Update}}: {{The}} {{{\textsc{ORCA}}}} {{Program System}}—{{Version}} 6.0. \emph{WIREs Computational Molecular Science} \textbf{2025}, \emph{15}, e70019\relax
\mciteBstWouldAddEndPuncttrue
\mciteSetBstMidEndSepPunct{\mcitedefaultmidpunct}
{\mcitedefaultendpunct}{\mcitedefaultseppunct}\relax
\EndOfBibitem
\bibitem[Schmerwitz \latin{et~al.}(2024)Schmerwitz, Ásgeirsson, and Jónsson]{schmerwitzImprovedInitializationOptimal2024}
Schmerwitz,~Y. L.~A.; Ásgeirsson,~V.; Jónsson,~H. Improved {{Initialization}} of {{Optimal Path Calculations Using Sequential Traversal}} over the {{Image-Dependent Pair Potential Surface}}. \emph{Journal of Chemical Theory and Computation} \textbf{2024}, \emph{20}, 155--163\relax
\mciteBstWouldAddEndPuncttrue
\mciteSetBstMidEndSepPunct{\mcitedefaultmidpunct}
{\mcitedefaultendpunct}{\mcitedefaultseppunct}\relax
\EndOfBibitem
\bibitem[Lee \latin{et~al.}(1988)Lee, Yang, and Parr]{leeDevelopmentColleSalvettiCorrelationenergy1988}
Lee,~C.; Yang,~W.; Parr,~R.~G. Development of the {{Colle-Salvetti}} Correlation-Energy Formula into a Functional of the Electron Density. \emph{Physical Review B} \textbf{1988}, \emph{37}, 785--789\relax
\mciteBstWouldAddEndPuncttrue
\mciteSetBstMidEndSepPunct{\mcitedefaultmidpunct}
{\mcitedefaultendpunct}{\mcitedefaultseppunct}\relax
\EndOfBibitem
\bibitem[Becke(1988)]{beckeDensityfunctionalExchangeenergyApproximation1988}
Becke,~A.~D. Density-Functional Exchange-Energy Approximation with Correct Asymptotic Behavior. \emph{Physical Review A} \textbf{1988}, \emph{38}, 3098--3100\relax
\mciteBstWouldAddEndPuncttrue
\mciteSetBstMidEndSepPunct{\mcitedefaultmidpunct}
{\mcitedefaultendpunct}{\mcitedefaultseppunct}\relax
\EndOfBibitem
\bibitem[Becke(1993)]{beckeDensityFunctionalThermochemistry1993}
Becke,~A.~D. Density‐functional Thermochemistry. {{III}}. {{The}} Role of Exact Exchange. \emph{The Journal of Chemical Physics} \textbf{1993}, \emph{98}, 5648--5652\relax
\mciteBstWouldAddEndPuncttrue
\mciteSetBstMidEndSepPunct{\mcitedefaultmidpunct}
{\mcitedefaultendpunct}{\mcitedefaultseppunct}\relax
\EndOfBibitem
\bibitem[Hehre \latin{et~al.}(2003)Hehre, Ditchfield, and Pople]{hehreSelfConsistentMolecular2003}
Hehre,~W.~J.; Ditchfield,~R.; Pople,~J.~A. Self—{{Consistent Molecular Orbital Methods}}. {{XII}}. {{Further Extensions}} of {{Gaussian}}—{{Type Basis Sets}} for {{Use}} in {{Molecular Orbital Studies}} of {{Organic Molecules}}. \emph{The Journal of Chemical Physics} \textbf{2003}, \emph{56}, 2257--2261\relax
\mciteBstWouldAddEndPuncttrue
\mciteSetBstMidEndSepPunct{\mcitedefaultmidpunct}
{\mcitedefaultendpunct}{\mcitedefaultseppunct}\relax
\EndOfBibitem
\bibitem[Weigend(2006)]{weigendAccurateCoulombfittingBasis2006}
Weigend,~F. Accurate {{Coulomb-fitting}} Basis Sets for {{H}} to {{Rn}}. \emph{Physical Chemistry Chemical Physics} \textbf{2006}, \emph{8}, 1057--1065\relax
\mciteBstWouldAddEndPuncttrue
\mciteSetBstMidEndSepPunct{\mcitedefaultmidpunct}
{\mcitedefaultendpunct}{\mcitedefaultseppunct}\relax
\EndOfBibitem
\bibitem[Wigner(1938)]{wignerTransitionStateMethod1938}
Wigner,~E. The Transition State Method. \emph{Transactions of the Faraday Society} \textbf{1938}, \emph{34}, 29--41\relax
\mciteBstWouldAddEndPuncttrue
\mciteSetBstMidEndSepPunct{\mcitedefaultmidpunct}
{\mcitedefaultendpunct}{\mcitedefaultseppunct}\relax
\EndOfBibitem
\bibitem[Vineyard(1957)]{vineyardFrequencyFactorsIsotope1957}
Vineyard,~G.~H. Frequency Factors and Isotope Effects in Solid State Rate Processes. \emph{Journal of Physics and Chemistry of Solids} \textbf{1957}, \emph{3}, 121--127\relax
\mciteBstWouldAddEndPuncttrue
\mciteSetBstMidEndSepPunct{\mcitedefaultmidpunct}
{\mcitedefaultendpunct}{\mcitedefaultseppunct}\relax
\EndOfBibitem
\end{mcitethebibliography}

\end{document}